\documentstyle[11pt,aasms4]{article}
\slugcomment{}
\lefthead{KUBOTA ET AL.}
\righthead{SPECTRAL TRANSITIONS OF TWO X-RAY SOURCES IN IC~342}
\slugcomment{}
\nonstopmode

\begin{document}

\title{DISCOVERY OF SPECTRAL TRANSITIONS FROM TWO \\
      ULTRA-LUMINOUS COMPACT X-RAY SOURCES IN IC~342} 
\author{A. {\sc Kubota},$^{1}$ T. {\sc Mizuno},$^{2}$ K. {\sc Makishima},$^{1}$  
Y. {\sc Fukazawa},$^{2}$\\ 
J. {\sc Kotoku},$^{1}$  T. {\sc Ohnishi},$^{1}$ and M. {\sc Tashiro}$^{1}$}
\affil{1:Department of Physics,  University of Tokyo, 
7-3-1 Hongo, Bunkyo-ku, Tokyo, Japan 113-0033}
\centerline{\it e-mail : aya@amalthea.phys.s.u-tokyo.ac.jp}
\affil{2:Department of Physical Science, Hiroshima University, 
1-3-1 Kagamiyama,\\ Higashi-Hiroshima, Japan 739-8526}

\begin{abstract}
Two {\it ASCA} observations were made of two 
ultra-luminous compact X-ray sources (ULXs), 
Source~1 and Source~2, in the spiral galaxy IC~342.
In the 1993 observation, 
Source~2 showed a 0.5--10 keV luminosity of 
$6 \times 10^{39}$ ergs s$^{-1}$ (assuming a distance of 4.0 Mpc),
and a hard power-law spectrum of photon index $\sim 1.4$.
As already reported,
Source~1 was $\sim 3$ times brighter on that occasion, 
and exhibited a soft spectrum represented by
a multi-color disk model of inner-disk temperature $ \sim 1.8$ keV.
The second observation made in February 2000 revealed 
that Source~1 had made a transition into a hard spectral state,
while Source~2 into a soft spectral state. 
The ULXs are therefore inferred to exhibit two distinct spectral states,
and sometimes make transitions between them.
These results significantly reinforce the scenario 
which describes ULXs as mass-accreting black holes.
\end{abstract}


\section{Introduction}
Arm regions of nearby spiral galaxies have long been known to harbour
luminous point-like X-ray sources (Fabbiano 1989, 1998; Read et al. 1997; Roberts \& Warwick 2000).
Except for several identified with young supernova remnants,
they remained unidentified in other frequencies.
Being often time variable,
these sources are thought to be accreting single objects;
we hence call them ultra-luminous compact X-ray sources 
(ULXs; Makishima et al. 2000, hereafter Paper~I).
With the X-ray luminosity reaching $10^{39-40}$ ergs s$^{-1}$, 
the ULXs are suspected to contain stellar black holes (BHs)
of mass up to $\sim 100~M_\odot$ or more 
(e.g., Colbert et al. 1995; Fabbiano 1998; Immler et al. 1999; Roberts \& Warwick 2000), 
so as not to violate the Eddington limit.
However, this much exceeds the mass estimates ($5-15~ M_\odot$; Tanaka \& Lewin 1995)
for accreting stellar BHs in the Galaxy and the Large Magellanic Cloud.
There is no consensus on the presence of $\sim 100~M_\odot$ BHs,
even though such medium-mass BHs have been proposed under slightly different context 
(Colbert \& Mushotzky 1999; Matsumoto \& Tsuru 1999; Ptak \& Griffiths 1999).
Furthermore, we do not find ULXs in the Milky Way or M31.
Therefore, the BH scenario of ULXs remained inconclusive,
and their nature has remained a big mystery.

A great leap has been achieved through spectroscopy of a dozen ULXs with {\it ASCA} 
(Takano et al. 1994; Okada et al. 1998; Colbert \& Mushotzky 1999; 
Mizuno et al. 1999; Mizuno 2000; Kotoku et al. 2000; Paper~I), 
as the spectra of majority of them have been reproduced by so-called 
multi-color disk model (MCD model; Mitsuda et al. 1984)
that describes emission from an optically-thick standard accretion disk
around a BH (Shakura \& Sunyaev 1973).
These results significantly reinforce the BH scenario of ULXs,
since the MCD model can explain the soft X-ray spectra of 
Galactic/Magellanic black-hole binaries (BHBs) in ``soft (or high) state''
(e.g., Makishima et al. 1986; Ebisawa et al. 1993;  Dotani et al. 1997; 
Kubota et al. 1998; Feroci et l. 1999; Paper~I).

The scenario is nevertheless still clouded by a self-inconsistency,
that the measured inner-disk temperature of ULXs, $T_{\rm in} = 1.0-1.8$ keV,
is too high for the implied high BH mass (Colbert \& Mushotzky 1999; Paper~I).
Although the scenario may be salvaged by invoking rapid BH rotation
(Mizuno et al. 1999; Paper~I; Mizuno 2000; also Zhang et al. 1997b)
and/or ``slim disk'' concept (Abramowicz et al. 1988; Watarai et al. 2000),
it is important to assemble further observational clues.

Here, we report further evidence supporting the BH interpretation of ULXs,
i.e., transitions of some ULXs between two distinct spectral states.
These results are based on two observations with {\it ASCA} (Tanaka et al. 1994) of two ULXs, 
both in the spiral galaxy IC~342 at a distance of 4.0 Mpc (Tully 1988).

\placefigure{fig:images}

\section{Observation}
The first observation of IC~342 was performed on 1993 September 19,
as reported by Okada et al. (1998).
The obtained X-ray image is reproduced in Figure~\ref{fig:images}a.
In addition to unresolved emission from the nuclear region,
we see two ULXs, the brighter Source~1 and the fainter Source~2,
both known from the {\it Einstein} era (Fabbiano \& Trinchieri 1987).
On this occasion, Source~1 exhibited clear short-term variability (Okada et al. 1998).

We observed IC~342 with {\it ASCA} again on 2000 February 24--March 1.
The GIS was operated in the normal PH mode,
while the SIS in the Faint mode. 
Although the acquired X-ray image, Figure~\ref{fig:images}b,
again reveals the two sources,
their relative intensities have reversed from those in 1993.
As quantified in Table~1,
the Source~1 flux decreased to $\sim 30\%$ of its 1993 value,
while that of Source~2 increased by a factor of 1.8 meantime.

\placetable{tbl:specfits}

\section{Data analysis and results}
\subsection{Data in 1993}
Spectral results on Source~1 from the 1993 observation are already 
described in Okada et al. (1998), Paper~I, and Mizuno (2000).
Referring to Paper~I, we simply quote here
that the 1993 spectra of Source~1 can be reproduced successfully
by an absorbed MCD model of $T_{\rm in} \sim 1.8$ keV.
In Table~1, we reproduce essence of these results.

We here analyze the GIS/SIS spectra of Source~2 after Mizuno (2000).
We selected good data in the same manner as in Okada et al. (1998),
and accumulated photons around the Source~2 centroid, separately for the four detectors.
We subtracted background using blank-sky data.
We then added the SIS0 and SIS1 spectra into a single SIS spectrum,
and those from GIS2 and GIS3 into a single GIS spectrum.
The net exposure became 38 ks for the GIS and 36 ks for the SIS.
The derived Source~2 spectra, shown in Figure~\ref{fig:specfits}a, 
are much harder than those of Source~1, 
and a joint GIS/SIS fit with an absorbed power-law model of 
photon index $\Gamma \sim 1.4$ has been successful (Table~1).
Although the absorbed MCD fit is also acceptable,
it requires an unrealistically high disk temperature of $T_{\rm in} \sim 3.0$ keV (Table 1).
An absorbed Bremsstrahlung model of temperature $>23$ keV is also acceptable ($\chi^2/\nu = 101.1/87$).
Considering typical spectra of BHBs,
we consider the power-law fit to be most appropriate for Source~2.

\placefigure{fig:specfits}

\subsection{Data in 2000}
We use only the GIS data for our spectral study of the 2000 observation,
because the long-term SIS degradation has made its low energy response rather uncertain. 
We selected good GIS data using criteria of geomagnetic cutoff rigidity $>6$ GeV, 
and the target elevation above the earth rim $> 5^\circ$.
This has yielded 276 ks of good exposure.

Through the long observation,
Source~1 and Source~2 both showed a relatively steady 0.7--10 keV intensity of 
0.027 c s$^{-1}$ and 0.045 c s$^{-1}$ per GIS detector respectively, 
with mild variation.
We here utilize the whole length of GIS data for spectral evaluation,
by accumulating events over two circular regions of radius 
$2.'5$ around Source~1 and $3.'0$ around Source~2.
By subtracting background utilizing blank-sky data,
and adding data from the two GIS detectors,
we have obtained the Source~1 and Source~2 spectra,
shown in Figures \ref{fig:specfits}b and \ref{fig:specfits}c, respectively.

The Source~1 spectrum has become much harder and less convex than in 1993,
and fitted approximately ($\chi^2/\nu = 119.8/88$) 
with an absorbed power-law of $\Gamma \sim 1.8$.
However at 7--10 keV, we observe negative residuals suggestive of an Fe-K edge.
By introducing an edge absorption at $\sim 8.4$ keV,
the fit became acceptable (Table~1);
an $F$-test indicates that the edge feature is significant at 99.5\% confidence.
Although the MCD fit is not too bad (Table~1),
the absorption it requires falls below the Galactic value of $\sim 3 \times 10^{21}$ cm$^{-2}$.
We therefore consider the power-law fit with an Fe-K edge appropriate.

In contrast, the Source~2 spectrum has become significantly more convex than in 1993.
It can now be expressed adequately with an MCD model of $T_{\rm in}=1.6$ keV,
whereas the power-law fit is unacceptable (Table 1).
Thus, the Source~2 spectrum in 2000 is close in shape to the Source~1 spectrum in 1993,
except for a higher photoelectric absorption.
An inclusion of a separate power-law component, with photon index fixed at 2.5, 
yields only an upper limit of 15\% in terms of its 0.5--10 keV flux,
and the MCD parameters are not affected beyond the fitting errors (see also \S~3.3 of Paper~I).
This indicates dominance of the optically-thick disk emission in the {\it ASCA} band.

To examine possible changes in the absorption,
we fitted the Source~1 spectra in 1993 and that in 2000 
with the MCD model and the power-law model respectively, under a common absorption.
We obtained an acceptable ($\chi_\nu^2=1.13$) joint fit, 
with $N_{\rm H} = 5.3 \pm 0.5$ cm$^{-2}$ and little changes in the other parameters.
Similarly, the 1993 power-law fit and the 2000 MCD fit to the Source~2 spectra
accepted ($\chi_\nu^2=0.94$) a common absorption of $N_{\rm H} = 18.1 \pm 0.8$ cm$^{-2}$.
Therefore, the transition of neither source was accompanied by measurable changes in the absorption.

\section{Discussion}
In the first {\it ASCA} observation of IC~342, 
Source~1 exhibited a high luminosity exceeding $10^{40}$ ergs s$^{-1}$ (Table~1),
prominent short-term variation,
and a convex spectrum expressed with an MCD model
of a high disk temperature, $T_{\rm in} \sim 1.8$ keV.
In Paper~1, these features have been identified as prototypical ULX properties.

In the same observation, Source~2 showed a distinct hard spectrum
expressed with a power-law of $\Gamma \sim 1.4$.
Among the ULXs so far studied with {\it ASCA}, another object, 
i.e., Source~A in NGC~1313 observed in 1993, exhibited a similarly hard spectrum 
with $\Gamma \sim 1.7$ (Petre et al. 1994; Colbert \& Mushotzky 1999; Mizuno 2000);
the remaining objects showed the MCD-type spectra (Paper~I).
We hence presume that there are two subtypes of ULXs,
a majority exhibiting the soft MCD-type spectra,
and a minority showing the hard power-law spectra.
However, from these results alone, we cannot tell 
whether or not these subtypes represent intrinsically distinct two source populations.

The second {\it ASCA} observation made in 2000 revealed
that Source~1 had made a soft-to-hard state transition, 
whereas Source~2 a hard-to-soft one,
although their reciprocal behavior must be a chance coincidence.
To visualize this, we present in Figure~\ref{fig:deconvolved} their deconvolved spectra.
These results immediately imply that;
(1) the two ULXs must be accreting compact objects,
to make such drastic spectral changes in 7 years;
(2) the two ULX subtypes represent two well-defined states
(soft state and hard state) of the same source population;
(3) some (if not all) ULXs make transitions between the two spectral states;
and 
(4) at least in the {\it ASCA} band, the source is less luminous while in the hard state.
These properties give a conclusive support to the BH interpretation of ULXs,
because they are just what characterize the known BHBs 
(e.g., Maejima et al. 1984; Tanaka \& Lewin 1995; Tanaka \& Shibazaki 1996; 
Zhang et al. 1997a; Wilms et al. 2000).
A similar (though less convincing) example may be the aforementioned Source~A in NGC~1313,
because its spectrum turned very soft in 1995
($\Gamma \sim 2.8$; Colbert \& Mushotzky 1999; Mizuno 2000).

\placefigure{fig:deconvolved}

Thus, the long-lasting puzzle of ULXs are being unveiled,
but a series of new questions are arising in turn.
Which is the right solution to the 
``too high $T_{\rm in}$''  problem?
Are the inferred high BH masses of ULXs consistent with the current 
understanding of stellar BH formation?
Why are ULXs absent in Our Galaxy and M31?
Do they form a population distinct from the Galactic/Magellanic BHBs?
In short, the ULX formation scenario is yet to be clarified.

The authors thank the {\it ASCA} team members.
The present work is supported in part by 
the Grant-in-Aid for Center-of-Excellence, No. 07CE2002, 
from Ministry of Education, Science, Sports and Culture of Japan.

\begin{table}[h]
\caption{The spectral parameters of IC~342 Source~1 and Source~2, with 90\% confidence limits.}
\begin{center}
\begin{small}
\begin{tabular}{lcccccccc}
\hline 
Epoch & $f_{\rm x}^{a)}$ & $L_{\rm x}^{b)}$ & 
                \multicolumn{3}{c}{Fit with the MCD model}&  
                \multicolumn{3}{c}{Fit with the power-law model} \\
      &         & &$T_{\rm in}$ (keV) &$N_{\rm H}^{c)}$& $\chi^2/\nu$ 
           &Photon index &$N_{\rm H}^{c)}$& $\chi^2/\nu$\\
\hline 
\hline 
Source 1 \\
~~1993 $^{d)}$&10.2 & 1.9 & $1.77 \pm 0.05$ & $4.7 \pm 0.3$  & 137.4/135 & 
                     $1.90 \pm 0.05$ & $9.3 \pm 0.6$  & 266.5/135\\
~~2000 & 3.1 & 0.6 & $2.06 \pm 0.08$ & $1.9 \pm 0.4$  & 123.9/89  &
               $1.73 \pm 0.06^{e)}$ & $6.4 \pm 0.7^{e)}$  & 101.1/86$^{e)}$\\
\hline 
Source 2 \\
~~1993 & 4.1 & 0.8 &$3.03 \pm 0.30$ & $9.9 \pm 0.9$  & 96.4/87  &
                    $1.39 \pm 0.10$ &$14.3 \pm 1.6$  & 102.8/87 \\
~~2000 & 7.2 & 1.4 &$1.62 \pm 0.04$ &$18.2 \pm 0.8$  & 90.5/89  &
                     2.48$^{f)}$     &  31.8$^{f)}$   & 232.0/89 \\
\hline 
\end{tabular}
\end{small}
\end{center}
\begin{footnotesize}
\begin{itemize}
\setlength{\baselineskip}{3mm}
\setlength{\itemsep}{-2mm}
 \item[$^{a)}$] The 0.5--10 keV source flux at the top of atmosphere,
          in unit of $10^{-12}$ ergs s$^{-1}$ cm$^{-2}$.
 \item[$^{b)}$] The absorption-uncorrected 0.5--10 keV luminosity 
          in $10^{40}$ ergs s$^{-1}$,
          assuming 4.0 Mpc distance and an isotropic emission.
 \item[$^{c)}$] Column density for absorption assuming solar abundances,
          in unit of $10^{21}$ cm$^{-2}$.
 \item[$^{d)}$] Results taken from Paper~I.
 \item[$^{e)}$] An ionized Fe-K edge (at $8.4 \pm 0.3$ keV) is applied, 
          with an optical depth of $0.9 \pm 0.5$.
 \item[$^{f)}$] Errors are not shown since the fit is highly unacceptable.
\end{itemize}
\end{footnotesize}
\label{tbl:specfits}
\end{table}

\clearpage

\vspace*{-3cm}
\epsscale{0.8}
\plotone{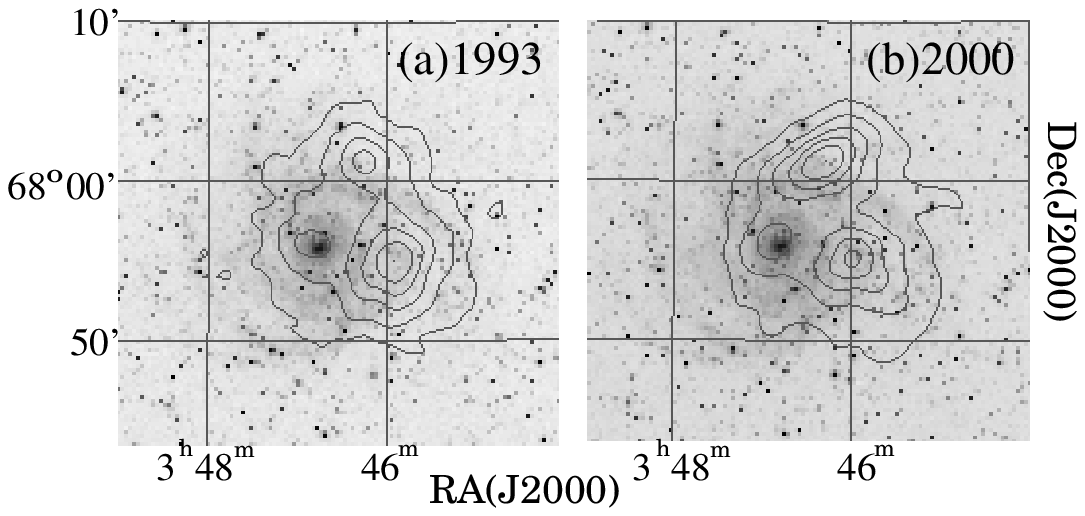}
\figcaption{
The 0.7--10 keV X-ray images of IC~342 taken with the {\it ASCA} GIS, 
shown after smoothing with a Gaussian kernel of $\sigma = 0'.5$
and superposed on the optical image.
The background has not been subtracted, 
and the X-ray contour levels are linear.
The source to the lower right of the optical nucleus is Source~1, 
while that to the upper right is Source~2.
(a) The data taken on 1993 September.
(b) Those of 2000 February.
\label{fig:images}
}

\vspace{1cm}
\epsscale{0.4}
\plotone{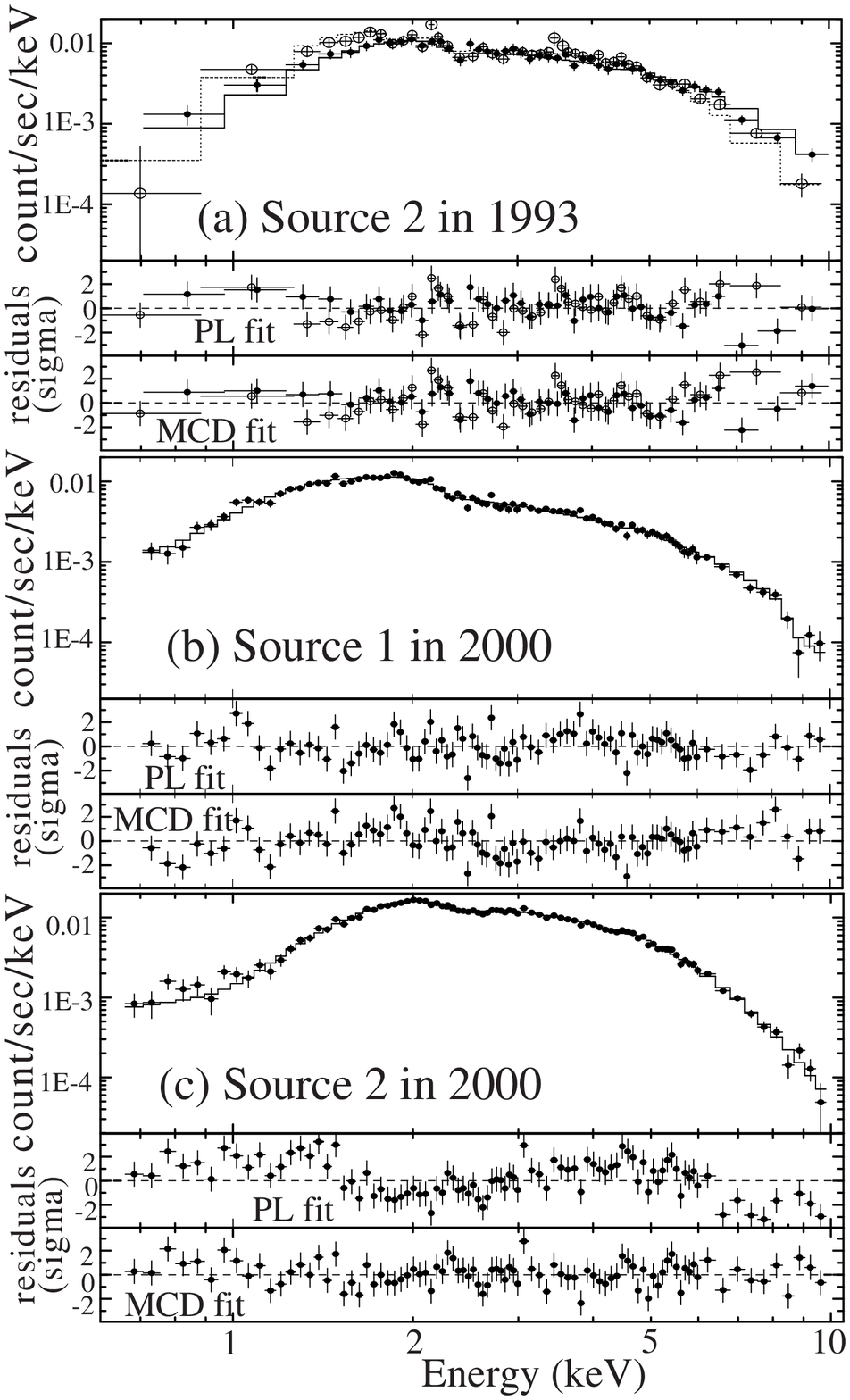}
\figcaption{
The {\it ASCA} spectra of IC~342 Source~1 and Source~2, 
together with prediction of the best-fit model (Table~1).
The model parameters are given in Table~1.
Fit residuals are shown for both the power-law and MCD models.
(a) The SIS (open circles) and GIS (filled circles) spectra of Source~2 in 1993, 
jointly fitted with an absorbed power-law model.
(b) The GIS spectrum of Source~1 in 2000,
fitted with an absorbed power-law model incorporating the ionized Fe-K edge.
(c) The GIS spectrum of Source~2 obtained in 2000,
fitted with an absorbed MCD model.
\label{fig:specfits}
}

\vspace*{1cm}
\epsscale{0.4}
\plotone{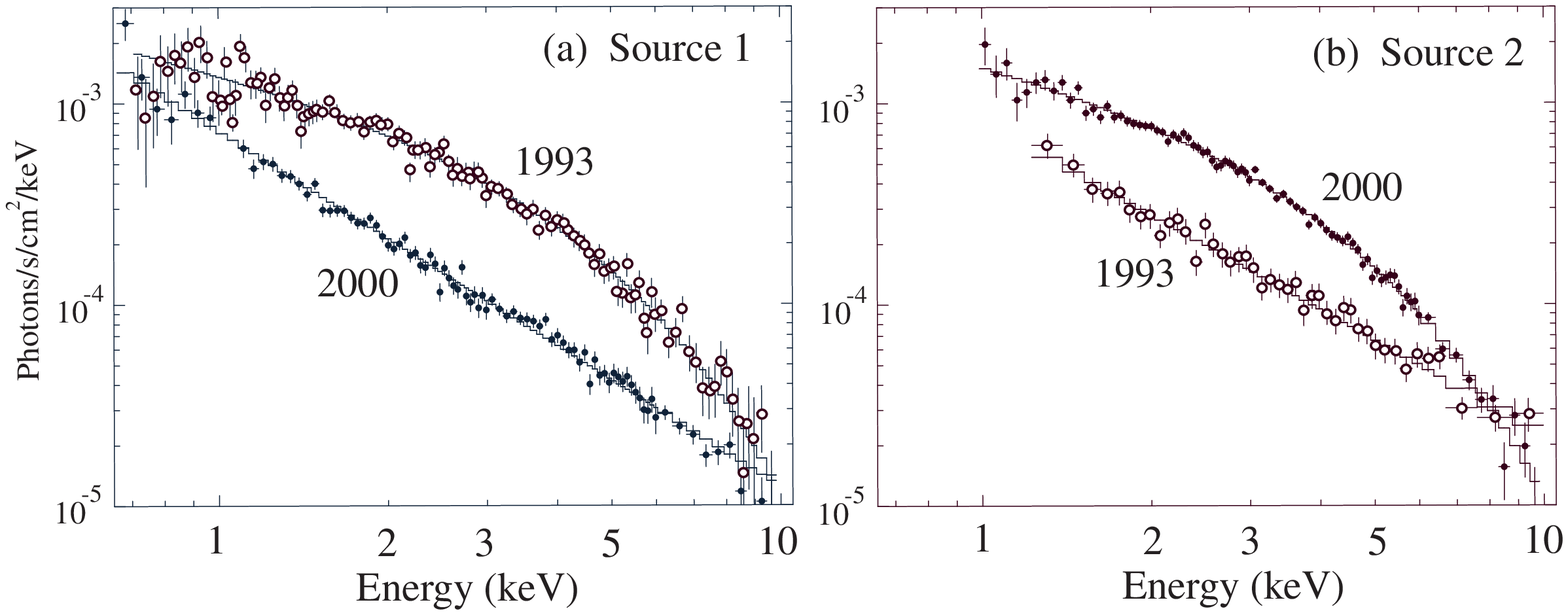}
\figcaption{
The {\it ASCA} GIS spectra of IC~342 
Source~1 (panel a) and Source~2 (panel b), 
presented after removing the instrumental responses and the photoelectric absorption.
Open and filled circles indicate the data taken in 1993 and 2000, respectively.
The Source~1 spectrum in 1993 and that of Source~2 in 2000 
are deconvolved using the MCD-model fits,
whereas the others using the power-law fits.
The solid lines indicate these best-fit models in their incident forms.
\label{fig:deconvolved}
}

\end{document}